\documentclass[runningheads]{llncs}
\usepackage{graphicx}
\usepackage{amssymb}
\usepackage{blindtext}
\usepackage{xcolor}
\usepackage{mathtools}

\def\attend #1{\textcolor{blue}{\textbf{#1}}\\}
\def\smalltitle #1{\vspace{10mm}\begin{center}{#1}\end{center}\vspace{10mm}}

\def\runo #1#2{#1_1, #1_2, ..., #1_{#2}}
\def\runz #1#2{#1_0, #1_1, ..., #1_{#2}}

\def\blank {blank}
\def\coaut #1{\textcolor{brown}{#1}}
\def\call {call number}

\DeclarePairedDelimiter{\ceil}{\lceil}{\rceil}

\begin{document}
\title{Read-once machines and the thermodynamic complexity of Maxwell's demons}
\titlerunning{Read-once machines and Maxwell's demons}
%
\author{Fırat Kıyak\inst{1} \and A. C. Cem Say\inst{2}}
\authorrunning{F. Kıyak, A. C. C. Say}
%
\institute{Department of Mathematics, Boğaziçi University, İstanbul, Turkey \and
Department of Computer Engineering, Boğaziçi University, İstanbul, Turkey}
\maketitle              
\begin{abstract}
The thermodynamical costs imposed by computational resource
limitations like memory and time have been investigated before. We focus on a new computational limitation, namely, the machine being allowed to scan  the input only once, and prove that it is associated with unavoidable thermodynamical cost, even in the presence of infinite time and memory resources. We identify this limitation to be the one suffered by Maxwell's demons. This provides a framework for  quantifying the complexity associated with an experiment that effectuates a  ``decrease'' in the entropy of a thermodynamic system.

\end{abstract}


\begin{section}{Introduction}

Research into the nature of computation as a physical process has revealed that the erasure or ``forgetting'' of information by computers (where the machine's logical state in the previous step is not determined uniquely by its present state) is associated with an unavoidable increase in entropy \cite{Lan61}, adding the concept of ``thermodynamical cost'' alongside the more familiar measures of computational complexity like time and memory. Bennett \cite{Ben73} showed that every computation that can be carried out by a standard Turing machine (TM) can also be performed by a reversible TM (i.e., one where no configuration has more than one predecessor) that leaves no ``garbage'' information (which would have to be erased sooner or later for reusing the computer)  in memory except the input and output strings at the completion of its execution. This result establishes that machines which possess the resources and flexibility to implement Bennett's construction in \cite{Ben73} can in principle operate with zero thermodynamical cost.

An interesting interplay between time, space and thermodynamic complexities is revealed when one considers restricting one's machines to be more efficient in terms of memory or time. ``Time-hungry'' vs. ``space-hungry'' reversible machines were demonstrated by Lange et al. \cite{LMT00} Recently, Y\i{}lmaz et al. \cite{YKUS22} considered the extremely constrained framework of real-time deterministic finite automata, and showed that different regular languages have different and unavoidable thermodynamic costs associated with their recognition by these machines.  

In this paper, we consider a limitation on computational models that, to our knowledge, has not been analyzed in the context of thermodynamical complexity until now. This is the ``read-once'' restriction; namely, a stipulation that the machine should read each symbol in its input string at most once. We show that machines limited in this manner are faced with unavoidable thermodynamical cost, even when they are allowed to use unbounded amounts of time and working memory. Although such machines can be ``wired'' so that their inter-configuration transitions are reversible, they have to leave garbage information in their memory at the end of their executions. We quantify the size of this garbage as a function of input length for a concrete (regular) language.

There is an interesting relationship between our results regarding the thermodynamic complexity of  read-once computation and the operation of Maxwell's demon \cite{LR02}, the celebrated thought experiment in which a ``nanorobot'' seems to violate the second law of thermodynamics by sorting gas molecules in different chambers of a container, seemingly decreasing the entropy of the system at no cost. Landauer's findings \cite{Lan61} are generally accepted to have resolved this question by noting that the demon has to erase its observational record when its memory is filled up, and therefore pay the erasure cost mentioned above. We  argue that the demon's problem is actually caused by the fact that it has read-once observational access to the situation inside the container.  (We show that ``read-twice'' machines with sufficiently large time and space bounds can always operate with zero thermodynamical cost.) We use our  result to demonstrate a way to quantify the  thermodynamic cost to be incurred for demons running specific procedures, and conclude with a discussion regarding demons and the second law in light of this approach.

\end{section}

\begin{section}{Preliminaries}
For concreteness, we start by describing a specialization of the well-known definition of Turing machines as a model of read-once computation. The reader will note that the proof of our main result (Theorem \ref{log cost input complexity}) can be adapted to other models (like random-access machines) if those are restricted to access the input in a single left-to-right scan as well.

A 
\textit{read-once Turing machine}
is a 6-tuple $(Q,\Sigma,\Phi,\Gamma,\delta,q_0)$, where
\begin{enumerate}
    \item $Q$ is the finite  set of states, such that $Q = B \cup S \cup \{q_{acc}, q_{rej}\}$ where 
    \begin{itemize}
        \item $B$ is the set of blind states,
        \item $S$ is the set of sighted states, such that $B \cap S = \emptyset$, and
        \item $q_{acc}$ and $q_{rej}$ are the accept and reject states,  respectively,
    \end{itemize}
    \item $\Sigma$ is the input alphabet, not containing the blank symbol  $\lhd$,
    \item $\Phi$ is the work tape alphabet,
    \item $\delta$ 
    is the transition function, described below, and
    \item  $q_0$ is the initial state, $q_0 \in Q$.
\end{enumerate}

$\Sigma_\lhd$ will be used to denote the union $\Sigma \cup \{\lhd\}$.

The transition function $\delta$ is constructed in two parts, as follows: For $q \in S$, $q' \in Q$, $\sigma \in \Sigma_\lhd$, $\phi, \phi' \in \Phi$,   and $d_w \in \{-1,0,+1\}$,  $\delta(q,\sigma,\phi)=(q',\phi',d_w)$ dictates that the machine will switch to state $q'$, write $\phi'$ on the work tape, move the input head one step to the right, and move the  work tape  head $d_w$ steps to the right, if it is originally in state $q$, scanning $\sigma$ and  $\phi$ on the two respective tapes. For $q\in B$, $\delta(q,\phi)=(q',\phi',d_w)$ dictates a transition that does not move the input head, is insensitive to the scanned input symbol, but is otherwise similar to the one described above regarding the changes to the state and the work tape.

A read-once TM halts with acceptance (rejection) when it executes a transition entering $q_{acc}$ ($q_{rej}$). Any transition that  moves the work tape head  off the left end of that tape leads to a rejection. The input string is assumed to be followed with infinitely many blank symbols on the input tape. 

As usual, such a machine is said to $recognize$ a language $L$ if it is guaranteed to eventually halt with acceptance whenever started with an input string in $L$, and will halt with rejection for all nonmembers of $L$. One can also  augment the model easily with an output tape and talk about the mapping from the set of input strings to the set of output strings as the function $computed$ by the machine. 

It is easy to see that a read-once Turing machine ``sees'' every symbol of its input string only once, and that every ``standard'' single-tape Turing machine $M$ recognizing some language  or computing a function can be converted to an ``equivalent'' read-once machine $M_{RO}$ that starts by copying the input string of length $n$ to its work tape (incurring a space overhead of $O(n)$) and simulates   $M$ thereafter. In the next section, we will show that the read-once restriction is also associated with unavoidable thermodynamical cost. 

\end{section}

\begin{section}{Read-once access and thermodynamical cost}

Consider a slightly less restricted model, where   the input string is allowed to be accessed not once, but twice. Such ``read-twice'' machines can compute any computable function reversibly. 

\begin{theorem}\label{two_access_reversible_possible}
For every standard one-tape Turing machine $M$, there exists a corresponding deterministic Turing machine $M_{RT}$ that scans its input string in only two  passes, computes the same function as $M$, and has zero thermodynamic cost.
\end{theorem}

\begin{proof}
The machine $M_{RT}$ uses its first pass over the input to copy it to work tape $T$. It then simulates $M$ reversibly by using $T$ as if it contains the input, e.g. by emulating Bennett's technique \cite{Ben73}. This procedure leaves no ``garbage'' except the  duplicate input copy on $T$ when it finishes. $M_{RT}$ then uses its second pass over the input to reversibly erase the contents of $T$, as described in \cite{Ben73}.
\end{proof}

Consider the language $L_a = \{w | w \text{ contains } a\}$ on the alphabet $\Sigma = \{a, b\}$. We will use $L_a$ to demonstrate that the read-once restriction imposes a thermodynamic cost even for machines with no bounds on memory and runtime.


\begin{theorem}\label{log cost input complexity}
Any read-once machine that recognizes  $L_a$ is associated with a thermodynamic cost of $\Omega(\log_2 n)$ bits for inputs of length $n$, regardless of its time or memory budgets. This bound is tight for machines with $\Omega(\log_2 n)$ bits of working memory.
\end{theorem}

\begin{proof}
 Let $M$ be some classical deterministic discrete machine (random access machine, Turing machine, etc.) which recognizes $L_a$ by performing a single left-to-right scan of the input string. Assume, without loss of generality, that $M$ has been structured to work ``reversibly'' (i.e. with every configuration having at most one predecessor) by keeping a record of its transition history as in Bennett's construction, so that it forgets nothing and therefore pays no thermodynamic cost until the end of its execution.\footnote{Landauer \cite{Lan61} showed that any irreversible operation performed by a machine 
necessitates a thermodynamical cost. If the atomic operations of our 
machine were irreversible, then  the 
thermodynamical cost would not be limited to the erasure of the final garbage. 
Therefore, we assume our machine works by reversible atomic operations. 
} Since $M$ is not allowed to perform the second input scan mentioned in Theorem \ref{two_access_reversible_possible},  its final configuration may include some garbage, whose eventual erasure would cause $M$ thermodynamic complexity. We will quantify the amount of this garbage.
 
 It is standard to use the word ``configuration'' to name the entire instantaneous description of such a machine. We will find it useful to distinguish the decription of the ``moving parts'' of $M$ from its fixed input string. For any configuration $c$ of $M$ on some input string $\boldsymbol{\sigma}$, we call the collection of all items in $c$ with the exception of $\boldsymbol{\sigma}$  the \textit{dynamic part} ($dp$) of $c$.\footnote{For a Turing machine, the $dp$ would consist of the internal state, the contents and head positions of the work tape(s), and the input head position.} Since a reversible machine is not supposed to forget its input string, the garbage of $M$  is only the $dp$ of its final configuration.

The fact that $M$ has no configurations with more than one predecessor has some interesting implications when one restricts attention to the $dp$'s and examines the ways in which the $dp$'s can transition to one another. For  any particular $dp$, say, $dp_x$,  call the  position of the input symbol that $M$ is currently ``seeing'' according to  $dp_x$ (represented by the numerical value of the  position of the input head in a TM) the \textit{focus} of $dp_x$. Since $M$'s transitions are reversible, we can conclude that:
\begin{itemize}
    \item No $dp$ can receive transitions from two $dp$'s with different foci. To see this, assume that $dp_x$ receives a transition from $dp_y$, whose focus is $i$, when the $i$'th input symbol is $\sigma_t$, and from $dp_z$, whose focus is $j$ ($j\neq i$) when the $j$'th input symbol is $\sigma_u$. Let the $i$'th and $j$'th symbols of the input string $\boldsymbol{\sigma}$  be  $\sigma_t$ and $\sigma_u$, respectively. But this would mean that the two configurations $(dp_y,\boldsymbol{\sigma})$ and $(dp_z,\boldsymbol{\sigma})$ both transition to $(dp_x,\boldsymbol{\sigma})$, contradicting reversibility.
    \item No $dp$ can receive transitions reading the same input symbol from two $dp$'s with the same focus, since this would again mean two configurations transitioning into the same configuration.
    \item A $dp$ that receives a transition from a $dp$ with a blind state can receive no transition from any other $dp$, since that also would necessarily violate reversibility.
\end{itemize}
This lets us conclude that the only condition where two distinct $dp$'s, say, $dp_y$ and  $dp_z$, can transition into a single $dp$ occurs when $dp_y$ and  $dp_z$  have the same  focus and  those transitions consume different input symbols. 


Consider the  set $A=\{bb...b, bb...ba, bb...baa, ..., baa...a, aaa...a$\} containing all strings of the form $b^*a^*$ of length $n$. We claim that those $n+1$ input strings leave $M$ in different final $dp$'s. This is easy to see for $b^n$, which is the only string in the collection leading to a rejecting configuration. For any two distinct members of $A$ containing one or more $a$'s, say,  $\boldsymbol{\sigma_i}=b^ia^{n-i}$ and $\boldsymbol{\sigma_j}=b^ja^{n-j}$  ($i<j$), assume that the computations (which we will name $C_i$ and $C_j$, respectively) of $M$ on these two strings proceed so that they end up with the same final $dp$. Let us examine these two computations ``in parallel'' by considering the $dp$'s they enter after consuming corresponding symbols of their inputs.


Immediately upon consuming the $(i+1)$'st input symbol, $C_i$ and $C_j$ enter  necessarily different $dp$'s. (This is because $M$ must accept $b^iab^{n-i-1}$ and reject $b^{n}$.) The same reasoning lets us see that the computations must be in different $dp$'s while consuming the later input symbols up to the $(j+1)$'st position. Upon consuming the $(j+1)$'st input symbol, the two computations can not ``merge'' into the same $dp$, since that position contains the same symbol in both   $\boldsymbol{\sigma_i}$ and $\boldsymbol{\sigma_j}$, and such mergings are impossible, as we saw above. This same condition forces us to conclude that no merge will occur during the consumption of the later symbols, which are identical in  $\boldsymbol{\sigma_i}$ and $\boldsymbol{\sigma_j}$. We conclude that $C_i$ and $C_j$ cannot end in the same $dp$.

We have shown that there are at least $n+1$ distinct ``final'' $dp$'s that $M$ may end up in while running on inputs of length $n$. By elementary information-theoretic considerations, one can conclude that even the most efficient ``garbage compression'' procedure that  $M$ can employ must leave a garbage of at least $\log_2(n+1)$ bits for at least some input strings of that length.


We have proved a lower bound for the thermodynamic complexity. An upper bound can be shown by describing an effective machine which scans the input for $a$'s and accepts immediately upon seeing the first $a$. For instance, consider the following read-once Turing machine $N$. $N$'s state set is just $\{q_0,q_{acc},q_{rej}\}$, where $q_0$ is a sighted state. $N$ does not use its work tape. At $q_0$, $N$ transitions to $q_{acc}$ if the input symbol is $a$, and  to $q_{rej}$ if it sees the blank symbol. If the input symbol is $b$, $N$ remains at $q_0$ while advancing the input head. Note that a configuration $c$ of $N$ cannot have more than one predecessor; for any $c$ that does have a predecessor $c'$, the internal state of $c'$ must  be $q_0$, and the input tape head must be positioned at the cell immediately to the  left of the head position of $c$. There can be only one such configuration. Therefore, $N$ is reversible. Since there are $n+1$ different possible final positions for the input head, the amount of information that has to be forgotten when the final configuration is  reset will be no more than $\ceil{\log_2(n+1)}$ bits. 
\end{proof}

\begin{section}{Discussion: Experiments as algorithms}

We have proved that read-once access to the input necessitates a thermodynamic cost for the computation using a mathematical analysis involving the configurations of the machine. We wish to illustrate a connection between read-once machines and ``entropy-reducing'' experiments conducted by reacting to a long sequence of observation results with the following example.



Consider the following ``experimental'' setup: We have a container containing $N$ molecules of some ideal gas. The container is divided into right and left subdivisions by an adjustable shutter. A device  will measure the pressures in  the two subdivisions for $T$ discrete time units,\footnote{We will describe a suitable value for $T$ shortly.} and  will close the shutter in case there is a big pressure difference between the subdivisions, say, when all of the  gas has accumulated in the left subdivision.

Let the symbol $a$ stand for ``all of the gas has accumulated in the left chamber'', and $b$ for ``there are gas molecules in the right chamber''. The whole experiment is determined by a string $\boldsymbol{\sigma} = \sigma_1\sigma_2...\sigma_{T}$, which is formed by encoding the state of the system at time step $t$ with $\sigma_t$. A read-once machine $M$ is given this input.
$M$'s 
configurations distinguish whether it has seen an $a$ in the past or not. The shutter operates so that it is closed when $M$'s configuration indicates that an $a$ has been seen, and is left open otherwise.

The outcome of this experiment depends on whether $\boldsymbol{\sigma}=b^{T}$ or not. If $\boldsymbol{\sigma}$ equals $b^{T}$, then there will be no change to the macrostate of the gas container. If it does not, then the shutter will close  the first time $M$ reads an $a$, and it will stay closed until the end. In this case, the gas will be locked in the left chamber, decreasing its entropy by $N$ bits, since we now know one additional bit of information about the location of each of the $N$ molecules.

Let us use this setup to attempt an alternative proof (``by physics'') of Theorem \ref{log cost input complexity}: By the second law of thermodynamics, the  decrease in the entropy of the gas container must be ``compensated'' by an increase in the entropy associated by the remainder of the system, which is seen to correspond closely to our read-once machine for $L_a$. We establish the connection as follows.

By the usual assumptions, the probability of observing all the molecules in the left subdivision at any time step $t$ is $\frac{1}{2^N}$, and if the time between the observations are long enough, such events can be treated as independent.
Let $p$ denote the probability that $\boldsymbol{\sigma}$  contains $a$. The final macrostate of the system is uncertain: If $\boldsymbol{\sigma}=b^{T}$, it is the same macrostate we started with, whose entropy we shall call $S$. If $\boldsymbol{\sigma}$ contains an $a$, the final state's entropy is $S-N$. 
Since these two cases occur with probabilities $1-p$ and $p$, respectively, 
the entropy of the final state is at most one more than
$(1-p)*S + p(S-N) = S - pN$. 
This means that the entropy of the container decreases by at least $pN$ bits.




To obtain a lower bound to this decrease in entropy as a function of the length of the input \`a la Theorem \ref{log cost input complexity}, we need to focus on a range for $T$ that reflects this relationship. 
If $T$ is too small, then $p$ will be too small; for example $T=1$ implies $p=\frac{1}{2^N}$ and $pN=\frac{N}{2^N}$. One wishes to see how high one's lower bound can get. If $T$ is too high, then  the contribution of $p$ will stagnate near $1$, leading  the thermodynamical cost to asymptotically approach a constant value, breaking the analogy we seek with computational complexity. A good choice is  $T=2^N$, which makes the distribution of the number of $a$'s in $\boldsymbol{\sigma}$ close to the Poisson distribution with $\lambda=1$ for large $N$ by the law of rare events.
This sets $p\approx 1 - \frac{1}{e}$, which is greater than $\frac{1}{2}$. In other words, the entropy of the gas container is lowered by $N$ bits with   probability at least $\frac{1}{2}$, therefore, the entropy change of the  container, $\Delta S$, is at least $\frac{N}{2} = \frac{\log_2 T}{2}$.
We apply the second law  to the closed  system consisting of the container plus our ``experimental apparatus'' $M$ to conclude  that the final entropy imbalance that has to be paid for by  $M$ is at least $\frac{\log_2 T}{2}$ for inputs of length $T$.\footnote{The reader can contrast the ``back-of-the-envelope'' nature of this second proof with the formal treatment in Theorem \ref{log cost input complexity}.}

This  lower bound to the thermodynamic cost of $M$ is not as tight as the one found in the previous section, but it reveals a connection between such experiments and  read-once complexity that we will explore further. Crucially, this is a two-way connection: The apparatus in the experiment described above is literally a read-once machine scanning the string of observations and changing its state to reflect when an $a$ has been observed, and Theorem \ref{log cost input complexity} quantifies the size of the garbage (and the entropy increase associated with its erasure) that this procedure will incur. Let us use this insight to comment on a point raised in the literature. 




Earlier attempts at ``exorcising'' Maxwell's demon (i.e. resolving the apparent violation of the second law of thermodynamics due to its operation) which focused on the thermodynamic costs of measurements \cite{Sc15}
, were criticized by Bennett \cite{bennett1987demons},  who claimed that the costs associated with the erasure of the garbage information is enough to save the second law. Bennett  justified this only for Szilard's engine \cite{bennett1982thermodynamics,5390026}. There are no reasons, a priori, to think that the erasure costs will be just enough to save the second law, and this must be subjected to a mathematical analysis. Therefore, Bennett's exorcism may be viewed as incomplete, due to the number of bits that needs to be deleted after the computation may not be enough to save the second law. Mathematical guarantees like Theorem \ref{log cost input complexity} about the quantity of  garbage bits produced by a Maxwell's demon  in its memory can be useful in proving that the erasure costs of the demon are enough to save the second law, leaving no need for the other costs, like the ones associated with measurements.

To give an example, consider a more general experimental setting to the one we considered above: A demon waits for a big fluctuation in a system in equilibrium (think of a gas container again) to lock it in a lower entropy state. Recall that such a machine can be seen as reading an input string $\boldsymbol{\sigma}$ over the alphabet $\Sigma = \{a,b\}$, where $a$ indicates that the measurement resulted with the wanted fluctuation. This  is a read-once machine (the observations arrive in time order, and cannot be ``replayed''), and the fact that it will close the shutter in case of a fluctuation means that it has the ability to recognize whether the symbol $a$ is present in the input $\boldsymbol{\sigma}$: this machine can be interpreted to "accept" $\boldsymbol{\sigma}$  if the system is locked in a low entropy state at the end of the experiment. We have shown in Theorem \ref{log cost input complexity} that the most effective machines doing this task create around $\log_2 T$ bits of garbage entropy where $T$ is the total number of measurements. 
Proving that the decrease in the entropy of the system acted upon according to the experimental procedure does not exceed the thermodynamic complexity of the experiment would strengthen Bennett's claim that all costs are due to erasure. On the other hand, if there is such an experiment in which the system's entropy can be decreased by an amount more than $\log_2 T$, then it must mean that the demon suffers from thermodynamical costs other than the erasure costs.
\color{black}

 Some philosophical difficulties arise when one asks the question ``What if I  conduct the experiment myself instead of programming a machine to do it?'' An experiment has an associated algorithm, and its observable results should not depend on the entity conducting it. This means that we would still observe the same decrease in the entropy of the container in  the scenarios described above. How can one apply a quantitative second law in such cases?  One may think of the  experimenter-container pair as a closed system, like we did with the machine-container system, but this seems to require quantifying the entropy increase in the body of the experimenter to reach a  result about the final state of the container. 
 Our ability to quantify the thermodynamical costs of an experiment is also helpful in this regard.
 
Imagine programming a machine to conduct the experiment, and isolating the machine-container system. Apply the second law of thermodynamics to conclude that the entropy of the final state of the machine-container system is at least as large as the initial state. The entropy of the final state of the machine-container system is at most the sum of the entropies of the machine and the experiment. Therefore, we have the inequality
\begin{equation}
    S_f + H_f \geq S_i,
\end{equation}
where $S_i$ and $S_f$ are the initial and final entropies of the container, and $H_f$ is the entropy of the machine at the end of the computation. One can also write

\begin{equation} \label{second law almost}
    \Delta S \geq -H_f
\end{equation}

Regardless of who or what conducts the experiment, the entropy change in the experiment, $\Delta S$, will be the same. We could have taken some other machine which has entropy  $H_f'$ at the end of the computation. Theorem \ref{second law almost} will be valid for all those, so we conclude that
\begin{equation}\label{second law}
    \Delta S \geq -H,
\end{equation}
where $H$ is the thermodynamical cost of the experiment, i.e. the infimum of the thermodynamical costs of machines that can conduct the experiment.  Every experiment can be described as an algorithm that can be run on a machine, and that algorithm will have a definitive objective thermodynamical cost, a number which can be found by mathematical methods, like we did in Theorem \ref{log cost input complexity}.

A human experimenter conducting this experiment will find the same  entropy change $\Delta S$ in the container; therefore Inequality \ref{second law} applies as a general law for all types of experiments, and it gives the quantitative result we have sought. The entropy of a system can decrease by at most the thermodynamical cost of the experiment acting on it, and that cost can be calculated by analyzing the experiment as an algorithm.


\end{section}










\end{section}

\section*{Acknowledgements}

This research was partially supported by Boğaziçi
University Research Fund Grant Number 22A01P1.

%
%
\bibliographystyle{splncs04}
\bibliography{references}

\end{document}